\documentclass[reprint,superscriptaddress,preprintnumbers,
groupedaddress,
nofootinbib,
 amsmath,amssymb,
 aps]{revtex4-1}
\usepackage{multirow}
\usepackage{upgreek}
\usepackage{tensor}
\usepackage{float}
\usepackage{latexsym}
\usepackage{graphicx}
\usepackage{amssymb}
\usepackage{amsmath}
\usepackage{amsfonts}
\usepackage[hidelinks]{hyperref}
\usepackage{color}
\usepackage{cool}
\usepackage{dcolumn}
\usepackage{slashed}
\usepackage{mathrsfs}

\begin{document}
\preprint{TUM-HEP-1350/21}

\title{Fractional dark energy: phantom behavior and negative absolute temperature}


\author{Ricardo G. Landim}\email{ricardo.landim@tum.de}

\affiliation{Technische Universit\"at M\"unchen, Physik-Department T70,\\ James-Franck-Stra\text{$\beta$}e 1, 85748 Garching, Germany}


\date{\today}

\begin{abstract}
The fractional dark energy (FDE) model describes the accelerated expansion of the Universe through a nonrelativistic gas of particles with a noncanonical kinetic term. This term is proportional to the absolute value of the three-momentum to the power of $3w$, where $w$ is simply the dark energy equation of state parameter, and the corresponding energy leads to an energy density that mimics the cosmological constant. In this paper we expand the fractional dark energy model considering a non-zero chemical potential and we show that it may thermodynamically describe a phantom regime. The \textit{Planck} constraints on the equation of state parameter put upper limits on the allowed value of the ratio of the chemical potential to the temperature. In the second part, we  investigate the system of fractional dark energy particles with negative absolute temperatures (NAT). NAT are possible in quantum systems and in cosmology, if there exists an upper bound on the energy. This maximum energy is one ingredient of the FDE model and indicates a connection between FDE and NAT, if FDE is composed of fermions. In this scenario, the equation of state parameter is equal to minus one 
and, using cosmological observations, we find that  the transition from positive to negative temperatures is allowed at any redshift larger than one.
  \end{abstract}

\maketitle
\section{Introduction}

Observations of type-Ia supernovae (SNe) indicate that the Universe is currently undergoing a phase of accelerated expansion \cite{perlmutter1999,reiss1998}. A fluid with negative pressure, dark energy (DE), causes this expansion, but its nature is one of the major challenges in modern cosmology. 
The simplest candidate for DE is a cosmological constant, in agreement with most of the cosmological observations \cite{Aghanim:2018eyx}. However, the existing tensions in the $\Lambda$CDM model, specially the one arising from the determination of the Hubble constant from cosmic microwave background (CMB) data \cite{Aghanim:2018eyx} and its local measurement trough Cepheids \cite{Riess:2020fzl}, summed
to the lack of well-motivated explanations for the origin of such a constant leads to alternative candidates for DE. Among a plethora of options, there are  scalar and vector fields \cite{peebles1988,ratra1988,Frieman1992,Frieman1995,Caldwell:1997ii,Padmanabhan:2002cp,Bagla:2002yn,ArmendarizPicon:2000dh,Brax1999,Copeland2000,Vagnozzi:2018jhn,Koivisto:2008xf,Bamba:2008ja,Emelyanov:2011ze,Emelyanov:2011wn,Emelyanov:2011kn,Kouwn:2015cdw,Landim:2015upa,Landim:2016dxh,Banerjee:2020xcn}, metastable DE \cite{Szydlowski:2017wlv,Stachowski:2016zpq,Stojkovic:2007dw,Greenwood:2008qp,Abdalla:2012ug,Shafieloo:2016bpk,Landim:2016isc, Landim:2017kyz,Landim:2017lyq}, holographic DE \cite{Hsu:2004ri,Li:2004rb,Pavon:2005yx,Wang:2005jx,Wang:2005pk,Wang:2005ph,Wang:2007ak,Landim:2015hqa,Li:2009bn,Li:2009zs,Li:2011sd,Saridakis:2017rdo,Mamon:2017crm,Mukherjee:2016lor,Feng:2016djj,Herrera:2016uci,Forte:2016ben}, interacting DE \cite{Wetterich:1994bg,Amendola:1999er,Guo:2004vg,Cai:2004dk,Guo:2004xx,Bi:2004ns,Gumjudpai:2005ry,Yin:2007vq,Costa:2013sva,Abdalla:2014cla,Costa:2014pba,Landim:2015poa,Landim:2015uda,Costa:2016tpb,Marcondes:2016reb,Landim:2016gpz,Wang:2016lxa,Farrar:2003uw,micheletti2009,Yang:2017yme,Marttens:2016cba,Yang:2017zjs,Costa:2018aoy,Yang:2018euj,Landim:2019lvl,Vagnozzi:2019kvw}, models using extra dimensions \cite{dvali2000}, etc.

Recently, an alternative explanation for the cosmological constant was proposed in \cite{Landim:2021www}, in the so-called fractional dark energy (FDE) model. In this setup, the energy of the system has a noncanonical kinetic term, proportional to $p^{3w}$, where $p$ is the three-momentum and $w$ is the DE equation of state parameter. The corresponding energy density is constant and mimics the one of the cosmological constant. 
Its smallness, however, is not a completely free parameter, but  may arise from a Fermi-Dirac integral and it is related to the particle's minimum energy. 
Furthermore, the energy eigenvalue is obtained from fractional quantum mechanics (see \cite{laskin2018fractional} for a recent review on fractional quantum mechanics), where the Laplacian has a fractional power (leading to the Riesz integral). In fact, fractional calculus has been used in different contexts, such as in fractional quantum mechanics \cite{laskin2000fractional,Laskin:2002zz,guo2006some,bayin2012consistency,bayin2012comment,dong2007some,de2010fractional},   Newtonian gravity \cite{Giusti:2020kcv,Giusti:2020rul} and quantum cosmology \cite{Moniz:2020emn,Rasouli:2021lgy}.

 In this paper we further investigate the FDE model in two contexts. First, we analyze the influence of a non-zero chemical potential. In this scenario, the equation of state can be smaller than minus one, indicating a phantom behavior 
 that comes only from the gas properties. Taking the measurements of $w$ from \textit{Planck} into account \cite{Aghanim:2018eyx} for the $w$CDM model, we are able to obtain upper limits for the ratio of the chemical potential to the temperature.
 
 Since the FDE model has a maximum energy as an ingredient to avoid a divergence when the momentum goes to zero, it is natural to investigate whether FDE particles can have negative absolute temperatures (NAT).  NAT were  initially predicted  and explored in the late 1940s and in the 1950s, both experimentally  (in a crystal) \cite{purcell1951nuclear}, and theoretically \cite{onsager1949statistical,Ramsey:1956zz}. They require an upper bound on the energy since the Boltzmann distribution function, for instance, would diverge if $T<0$ and the energy is unlimited \cite{Ramsey:1956zz,klein1956negative}. A revival of NAT occurred after the experimental realization of NAT for motional degrees of freedom \cite{braun2013negative}, rather than the previously done localized spin systems \cite{purcell1951nuclear,oja1997nuclear,medley2011spin}. In cosmology, an interesting consequence of negative temperatures is negative pressure, thus indicating a possible relation between DE and NAT. In \cite{Vieira:2016lyj}, NAT in cosmology were further investigated. Hence, we explore the connection between NAT and the FDE model in the second part of this work 
 and we use cosmological observations to investigate the transition from positive temperature to NAT.

This paper is organized in the following manner. In section \ref{sec:fde} we review the FDE model. Section \ref{sec:chemical} is devoted to analyzing the system of FDE particles when the chemical potential is non-zero, while in section \ref{sec:NAT} we investigate the connection between FDE and NAT. In section \ref{sec:conclu} we summarize the work and present our conclusions. We will use natural units $\hslash=c=k_B=1$ throughout the text, unless explicitly stated.

    \section{Fractional dark energy}\label{sec:fde}

\sloppy In this section we review the main features of the FDE model presented in \cite{Landim:2021www}. FDE is composed of particles that satisfy the fractional Schr\"odinger equation.   In the context of fractional calculus a fractional derivative can be defined as the Riemann-Liouville derivative  \cite{oldham2006fractional}
\begin{equation}
    _aD^\alpha_xf(x)=\frac{1}{\Gamma(n+1-\alpha)}\frac{d^{n+1}}{dx^{n+1}}\int^x_a(x-y)^{n-\alpha}f(y)dy\,, 
\end{equation}
for $n\leq \alpha<n+1$, where $\Gamma(n+1-\alpha)$ is the Gamma function. The inverse operator is  the Riemann-Liouville fractional integration
operator
\begin{equation}
    _aD^{-\alpha}_xf(x)=\frac{1}{\Gamma(\alpha)}\int^x_a(x-y)^{\alpha-1}f(y)dy\,, \quad  \alpha>0\,,
\end{equation}
such that  $_aD^\alpha_b(_aD^{-\alpha}_bf(x))=f(x)$, with the operators satisfying the property $_aD^{\pm\alpha}_b(_aD^{\pm\beta}_bf(x))= {}_aD^{\alpha\pm\beta}_bf(x)$.

The operator $(\hslash \nabla)^\alpha$ is thus defined as \cite{laskin2018fractional}
\begin{equation}
    (\hslash \nabla)^\alpha= -\frac{\hslash^\alpha}{2\cos(\pi \alpha/2)}\Big[_{-\infty}D^{\alpha}_x +_{x}D^{\alpha}_\infty \Big]\,,
\end{equation}
 where $\alpha$ is the L\'evy index  $1<\alpha\leq 2$.

 The quantum Riesz fractional derivative (or integral, if $\alpha<0$) gives the fractional Laplacian operator 
 \begin{equation}\label{eq:inverse_op}
    (-\hslash^2\Delta)^{\alpha/2}\psi(\mathbf{r},t)= \frac{1}{(2\pi \hslash)^3}\int d^3p e^{i \mathbf{p}\cdot \mathbf{r}/\hslash}|\mathbf{p}|^{\alpha} \varphi(\mathbf{p},t)\,.
    \end{equation}
    Therefore, the fractional Schr\"odinger equation for FDE is
  \begin{equation}
    i\hslash\frac{\partial \psi(\mathbf{r},t) }{\partial t}=C (-\hslash^2\Delta)^{3w/2}\psi(\mathbf{r},t)\,,
\end{equation}
where $C$ is a constant with units of $[C]= \text{erg}^{1-3w}$cm$^{3w}$s$^{-3w}$ (or [energy]$^{1-3w}$, in natural units). 

The eigenvalue of the Hamiltonian operator gives a noncanonical kinetic term, which can be inserted into the nonrelativistic limit of the energy-momentum relation
\begin{equation}\label{eq:EMrelation2}
       \epsilon\approx m+\frac{p^2}{2 m}+\frac{C}{p^{-3 w}}\,,
   \end{equation}
   where $p\equiv |\mathbf{p}|$. When the non-relativistic particles cooled down below the rest mass ($C p^{3 w}>m$) the DE behavior started dominating.
   
   The DE number density and energy density are then respectively given by
   \begin{align}
    n=&-\frac{C^{-\frac{1}{w}}g}{6\pi^2w}\int_{\epsilon_{\min}}^{\epsilon_{\max}} \frac{\epsilon^{\frac{1}{w}-1} }{e^{\beta \epsilon}\pm 1}d\epsilon\,,\\
    =& -\frac{C^{-\frac{1}{w}}g}{6\pi^2w}\beta^{-\frac{1}{w}} \mathscr{F}^{u_{\max}}_{u_{\min}, \frac{1}{w}-1}\label{eq:n_de}\,,
\end{align}
   \begin{align}
    \rho= &-\frac{C^{-\frac{1}{w}}g}{6\pi^2w}\int_{\epsilon_{\min}}^{\epsilon_{\max}} \frac{\epsilon^{\frac{1}{w}} }{e^{\beta \epsilon}\pm 1}d\epsilon\,\\
    = & -\frac{C^{-\frac{1}{w}}g}{6\pi^2w}\beta^{-\frac{1+w}{w}} \mathscr{F}^{u_{\max}}_{u_{\min}, \frac{1}{w}}\,,\label{eq:rho_final}
\end{align}
where $g=2s+1$ is the spin multiplicity, $\beta=T^{-1}$, $u\equiv\beta \epsilon$, $\epsilon_{\min}\approx m$ is the nonrelativistic energy and $\epsilon_{\max}=\Lambda$ is a cutoff scale to avoid a divergence on the energy (\ref{eq:EMrelation2}). The  integral in the equation above  is
\begin{equation}
  \mathscr{F}^{u_{\max}}_{u_{\min,\frac{1}{w}}}\equiv \int_{u_{\min}}^{u_{\max}}\frac{u^{\frac{1}{w}} }{e^u\pm 1}du\,.
\end{equation}
The change of variables does not influence the limits of integration because the cutoff energy is sufficiently large and the nonrelativistic energy may be turned into $m_0+c_0T$, where $m_0$ is the particle's rest mass and $c_0$ is a constant of order of $m_0/T\sim 3$. This effective mass may resemble the analogue effective mass in solid-state physics and a further study of such a temperature dependence is subject of a future work. Depending on the value of $u_{\min}=\epsilon_{\min}/T$ the integral can be sufficiently small and $C$ can be of order of unity to give the correct observed vacuum energy. For example, if  $u_{\min}=10$ then $C=10^{-40}$ GeV$^4$, or $C=1$ GeV$^4$ for $u_{\min}=100$. Even for the apparent small value of $C$, we may notice that if it is proportional to the inverse of the Planck mass $C=\lambda^6/M_{Pl}^2$, then the new constant can be around $\lambda=10^{-2/6}$ GeV. In the context of fractional quantum field theory, the constant $C^{-1/(1-3w)}$ would be a length scale \cite{Calcagni:2021ipd, Calcagni:2021ljs,Calcagni:2021aap}.

\section{Fractional dark energy with non-zero chemical potential}\label{sec:chemical}

The discussion presented in the last section assumed a null chemical potential. If the chemical potential $\mu$ is now different from zero, it may give rise to a phantom behavior. 

The second law of thermodynamics can be written as \cite{Weinberg:1971mx}
\begin{equation}
  n T  d\sigma = d\rho -\frac{\rho+P}{n}dn\,,
\end{equation}
where  $\sigma$ is the entropy density per number density and $P$ is the pressure. Using the the particle number conservation, the continuity equation and the fact that $d\sigma$ is an exact differential, the equation above can be written as \cite{Silva:2002fi}
\begin{equation}\label{eq:Tdot}
   \frac{\dot{T}}{T}= \frac{\partial P}{\partial \rho}\frac{\dot{n}}{n}\,,
\end{equation}
which yields $n\propto T^{1/w}$ and $T\propto V^{-w}$.

Taking the Euler equation \cite{callen1998thermodynamics}
\begin{equation}
 Ts= \rho + P  -\mu n=(1+w)\rho-\mu n\,, 
\end{equation}
we can see that the entropy is positive even for $w<-1$, provided that $\mu <0$. The condition of positive entropy leads to a lower limit on the equation of state parameter \cite{Lima:2008uk}
\begin{equation}\label{eq:wmin}
    w\geq -1 +\frac{\mu n}{\rho}\,.
\end{equation}
Given the relations $n=n_0 a^{-3}$, $\rho=\rho_0 a^{-3(1+w)}$, $T=T_0a^{-3w}$ and $s=s_0a^{-3}$ one can obtain \cite{Lima:2008uk}
\begin{equation}
    \mu = \mu_0 a^{-3w}=\mu_0\frac{T}{T_0}\,,
\end{equation}
where
\begin{equation}
    \mu_0=\frac{1}{n_0}[-T_0s_0+(1+w)\rho_0]\,.
\end{equation}
The chemical potential is thus negative for a phantom behavior. The dependence of $\mu$ on the scale factor shows that the combination $\mu n/\rho$ in Eq. (\ref{eq:wmin}) is independent of the scale factor, i.e., $\mu n/\rho=\mu_0 n_0/\rho_0$.

A non-zero chemical potential modifies the DE number density and energy density to
\begin{align}
    n &= -\frac{C^{-\frac{1}{w}}g}{6\pi^2w}\beta^{-\frac{1}{w}} \mathscr{F}^{u_{\max},\, \beta\mu}_{u_{\min}, \frac{1}{w}-1}\label{eq:n_de2}\,,
\end{align}
   \begin{align}
    \rho &= -\frac{C^{-\frac{1}{w}}g}{6\pi^2w}\beta^{-\frac{1+w}{w}} \mathscr{F}^{u_{\max},\, \beta\mu}_{u_{\min}, \frac{1}{w}}\,,\label{eq:rho_final2}
\end{align}
where  here $u\equiv \beta \epsilon-\beta\mu$ and
\begin{equation}
  \mathscr{F}^{u_{\max,\,\beta\mu}}_{u_{\min,a}}\equiv \int_{u_{\min}}^{u_{\max}}\frac{(u+\beta\mu)^{a} }{e^u\pm 1}du\,.
\end{equation}
The result of the integration above  now depends on $\beta \mu$. Similarly to the case of null chemical potential we have the relation between the number density and energy density
\begin{equation}\label{eq:integra_u}
    \rho=\beta^{-1}\frac{\mathscr{F}^{u_{\max},\,\beta\mu}_{u_{\min},\frac{1}{w}}}{\mathscr{F}^{u_{\max},\,\beta\mu}_{u_{\min},\frac{1}{w}-1}}n\,.
\end{equation}

\begin{figure*}
    \centering
    \includegraphics[scale=0.55]{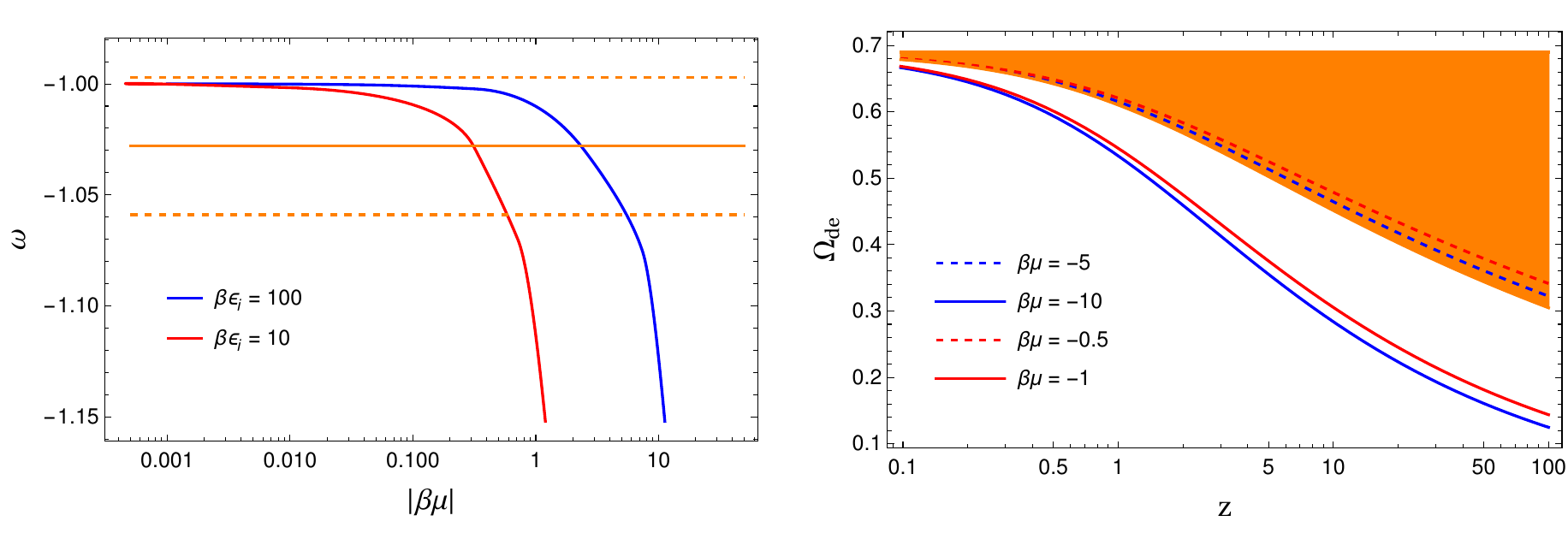}
    \caption{\textit{Left:} Equation of state parameter (\ref{eq:deltaw}) for different initial energies $\beta\epsilon_i$, as a function of $\beta\mu$. The orange lines represent the 68\% C.L.  constraints  on the equation of state parameter for the $w$CDM model ($w=1.028\pm 0.031$ \cite{Aghanim:2018eyx}), using the combination of \textit{Planck}, baryon acoustic oscillations (BAO) and SNe data. \textit{Right:} Redshift evolution of the DE density parameter using the equation of state on the left plot, for the two corresponding choices $\beta \epsilon_i=100$ (blue) and  $\beta \epsilon_i=10$ (red). The orange region represents the 68\% C.L. constrains from \textit{Planck}+BAO+SNe, for the phantom behavior. 
    }
    \label{fig:deltaw}
\end{figure*}

Using Eqs. (\ref{eq:n_de2}) and (\ref{eq:rho_final2}) the equation of state (\ref{eq:wmin}) becomes
\begin{equation}\label{eq:deltaw}
 w= -1 - \frac{|\mu_0|n_0}{\rho_0}=-1 -|\beta\mu|\frac{\mathscr{F}^{u_{\max},\,\beta\mu}_{u_{\min},\frac{1}{w}-1}}{\mathscr{F}^{u_{\max},\,\beta\mu}_{u_{\min},\frac{1}{w}}}\,.
\end{equation}
Notice that the combination $\beta\mu$ is independent of the scale factor. 
Eq. (\ref{eq:deltaw}) is constant and it may represent the $w$CDM model, in the phantom regime. Therefore we can compare the  equation of state for FDE for different values of $\beta \mu$, given by Eq. (\ref{eq:deltaw}), with the corresponding one for the $w$CDM model compatible with cosmological observations. Fig. \ref{fig:deltaw} presents this comparison for two different initial energies ($u_{\min}$) and the evolution of the DE density parameter. As it can seen, $\beta\mu$ should be roughly less than 5\% of $\beta \epsilon_i$ to be in agreement with the observations.

As it was shown in \cite{Landim:2021www} both energies can give the correct observed vacuum energy, provided that the constant $C$ is chosen accordingly.

\section{Negative absolute temperature}\label{sec:NAT}
The FDE model has an intrinsic cutoff to avoid a divergence in the energy and the minimum energy is the particle  mass.  These limits are similar to the ones in a scenario where a cosmological fluid possesses NAT, as investigated in \cite{Vieira:2016lyj}.

The canonical absolute temperature can be defined as the variation of the entropy with the internal energy at thermal equilibrium
\begin{equation}
    \frac{1}{T}=\left(\frac{\partial S}{\partial U}\right)_{V,N}\,,
\end{equation}
which can be negative.

 Necessary requirements for NAT are thermal equilibrium among the elements in the system and upper limit on the energy of the allowed states \cite{Ramsey:1956zz}. Classically a system is not found with NAT because there is no upper bound on its energy. However, quantum mechanical systems can be constructed such that they possess an energy upper bound. In order scenarios this upper bound can be seen as an energy cutoff \cite{Vieira:2016lyj}.

A system with NAT is hotter than the one with positive temperature because the former has more energy stored in it and can transfer its energy to any system at positive temperature it is in contact with \cite{purcell1951nuclear}.  Since a fluid with negative equation of state  will  have always an increasing temperature, it is natural to think that the NAT will be reached sometime in the evolution of this fluid. 

In a cosmological context a fluid with positive temperature has the following number density, energy density and pressure \cite{Vieira:2016lyj}
\begin{align}
n(T,\mu)  & = \int_m^{\Lambda} D(\epsilon) \mathcal{N}(T,\epsilon,\mu)d\epsilon\,,\\
\rho(T,\mu)  &  = \int_m^{\Lambda} \epsilon D(\epsilon) \mathcal{N}(T,\epsilon,\mu)d\epsilon\,,\\
P(T,\mu) &  = \beta^{-1}\int_m^{\Lambda} \epsilon D(\epsilon)\ln(1+e^{-\beta( \epsilon-\mu)})d\epsilon\,,\label{eq:pressure}
\end{align}
where $D(\epsilon)$ is the density of states and $\mathcal{N}(T,\epsilon,\mu)$ is the Fermi-Dirac distribution. The energy of the system is not bounded from above for bosons with no number conservation, therefore if $\mu=0$ a cosmological fluid with NAT can only be composed of fermions \cite{Vieira:2016lyj}. 

The symmetry of the Fermi-Dirac distribution and the logarithm in Eq. (\ref{eq:pressure}) allows the following relation between positive and negative temperatures
\begin{align}
    \mathcal{N}(T,\epsilon,\mu)&=\frac{1}{e^{\beta( \epsilon-\mu)}+1}=1-\frac{1}{e^{-\beta( \epsilon-\mu)}+1}\,,\nonumber\\
    &= 1- \mathcal{N}(-T,\epsilon,\mu)\,,\label{eq:N-T}\\
\ln\big[1+e^{-\beta (\epsilon-\mu)}\big]&= -\beta( \epsilon-\mu)+\ln\big[1+e^{\beta( \epsilon-\mu)}\big]\,.\label{eq:lnt-t}
\end{align}
As pointed out in \cite{Vieira:2016lyj}, Eq. (\ref{eq:N-T}) indicates the idea of ``holes'' and ``particles'', where holes are particles in the highest energy states (with $T<0$), which simply represent the absence of particles in states with positive temperature. The highest number density and energy density is thus when $\beta \rightarrow -\infty$, yielding
\begin{align}
    n_{\max}&= \int_m^{\Lambda} D(\epsilon) d\epsilon=\frac{C^{-\frac{1}{w}}g}{6\pi^2}\big(m_0^{\frac{1}{w}}-\Lambda^{\frac{1}{w}}\big)\,,\label{eq:nmax}\\
\rho_{\max} &  = \int_m^{\Lambda} \epsilon D(\epsilon) d\epsilon\stackrel{w=-1}{=}\frac{Cg}{6\pi^2}\big[\ln (\Lambda)-\ln(m_0)\big]\,,
\end{align}
\sloppy where the second equality in both equations comes from the fact that for FDE the density of states is $D(\epsilon)=-C^{-\frac{1}{w}}g/(6\pi^2w)\epsilon^{\frac{1}{w}-1}$. The lower limit of integration is $m=m_0$ because for negative temperatures $m=m_0+c_0|T|$ and $T=0^-$. From Eqs. (\ref{eq:N-T}) and (\ref{eq:nmax}) one can see that the maximum energy density and number density would be negative and thus unphysical, if bosons were considered instead of fermions. 

Using the properties in Eqs. (\ref{eq:N-T}) and (\ref{eq:lnt-t}) one can obtain the following relations
\begin{align}
    n(T,\mu)&=n_{\max}-n(-T,\mu)\,,\label{eq:nat}\\
    \rho(T,\mu)&= \rho_{\max} - \rho(-T,\mu)\,,\label{eq:rat}\\
    P(T,\mu)&= - \rho_{\max}+\mu n_{\max} -P(-T,\mu)\,.\label{eq:pat}
\end{align}

In our case, both particles and holes (i.e., particles with $T>0$ or $T<0$) will be described by a fluid with the same negative equation of state parameter, since the corresponding density of states arises from the noncanonical kinetic term (\ref{eq:EMrelation2}) in both regimes.

The number density and energy density for holes are still given by Eqs. (\ref{eq:n_de}) and (\ref{eq:rho_final}) (or Eqs. (\ref{eq:n_de2})   and (\ref{eq:rho_final2}) if $\mu\neq 0$), respectively, with the simple replacement $\beta\rightarrow |\beta|$ and the integrals are performed for negative $\beta$:
\begin{align}
    n_h &= -\frac{C^{-\frac{1}{w}}g}{6\pi^2w}|\beta|^{-\frac{1}{w}} \mathscr{F}^{u_{\max},\, \beta\mu}_{u_{\min}, \frac{1}{w}-1}\label{eq:n_de3}\,,\\
    \rho_h &= -\frac{C^{-\frac{1}{w}}g}{6\pi^2w}|\beta|^{-\frac{1+w}{w}} \mathscr{F}^{u_{\max},\, \beta\mu}_{u_{\min}, \frac{1}{w}}\,,\label{eq:rho_final3}
\end{align}
where here $u\equiv -|\beta|(\epsilon-\mu)$ and the subscripts ``p'' and ``h'' will indicate particles with $T>0$ and with $T<0$, respectively. Contrary to the case of positive $T$ ($\infty<\beta<0^+$), the negative temperature makes the number density to grow with $|\beta|$, because an increase in a negative temperature corresponds to an increase in $|\beta|$.\footnote{The temperature increases as $0^+,\dots, +\infty,\dots,-\infty, \dots, 0^-$ or in terms of $\beta$ as $+\infty,\dots, 0, \dots, -\infty$, which makes the description of NAT through $\beta $ more intuitive.}  The following relations also hold
\begin{align}
    \rho_h=|\beta|^{-1}\frac{\mathscr{F}^{u_{\max},\,\beta\mu}_{u_{\min},\frac{1}{w}}}{\mathscr{F}^{u_{\max},\,\beta\mu}_{u_{\min},\frac{1}{w}-1}}n_h\,.\label{eq:integra_u_minus_T}\\
   \frac{\dot{T}}{T}=-\frac{\dot{\beta}}{\beta}= \frac{\partial P}{\partial \rho}\frac{\dot{n}_{p,h}}{n_{p,h}}\,.\label{eq:Tdot_hp}
\end{align}
Eq. (\ref{eq:Tdot_hp})  also yields $n_p\propto \beta^{-\frac{1}{w}}$ and $n_h\propto |\beta|^{-\frac{1}{w}}$, thus the number density of holes increases up to $n_{\max}$.

The number conservation equation and the continuity equation for particles and holes are
\begin{align}
     \dot{n}_p+3Hn_p&=0\,,\label{eq:eq_n_p}\\
      \dot{n}_h+3H(n_h-n_{\max})&=0\,,\label{eq:eq_n_h}\\
      \dot{\rho}_p+3H(1+w)\rho_p&=0\,,\label{eq:eq_rho_p}\\
       \dot{\rho}_h+3H(1+w)\rho_h&=0\,.\label{eq:eq_rho_h}
\end{align}
The solution of Eq. (\ref{eq:eq_n_h}) is $n_h=n_{\max}-n_{\max}(a_*/a)^3$, where the scale factor $a_*$ represents the time when the initial abundance of particles was equal to the maximum number density $n_p^*=n_{\max}$. This solution means that in the early Universe all states with positive temperature were filled and $n_h=0$. As the Universe evolves $n_p$ decreases and $n_h$ increases up to $n_{\max}$. When $\beta\rightarrow 0$ both number densities are equal to half of the maximum number density, $n_p=n_h=n_{\max}$/2.  Since $n_h\propto |\beta|$  NAT increase as $|\beta|\propto n_{\max}(1-(a_*/a)^3)$,\footnote{Note that this evolution   is valid for $|\beta|<\infty$, because the number density for holes  is no longer given  by Eq. (\ref{eq:eq_rho_h2}) when $\beta\rightarrow -\infty$, but by Eq. (\ref{eq:nmax}).} while for positive temperatures the scaling is linear $T=\beta^{-1}\propto V$.

The number conservation  was required in \cite{Vieira:2016lyj} because for canonical particles the number conservation equation and the continuity equation gave inconsistent results. For example,  for nonrelativistic particles ($w=0$) Eq. (\ref{eq:eq_n_h}) can be multiplied by $m$, but it does not yield Eq. (\ref{eq:eq_rho_h}). However, this problem is solved for FDE. Taking the time derivative of Eqs. (\ref{eq:integra_u}) and (\ref{eq:integra_u_minus_T}), for particles and holes, respectively, and using Eqs. (\ref{eq:Tdot_hp}), (\ref{eq:eq_n_p}) and (\ref{eq:eq_n_h}), we obtain
\begin{align}
          \dot{\rho}_p&=-3H(1+w)\rho_p\,,\label{eq:eq_rho_p2}\\
       \dot{\rho}_h&=-3H(1+w)\rho_h+3H(1+w)\rho_h\frac{n_{\max}}{n_h}\,.\label{eq:eq_rho_h2}
\end{align}

Eq. (\ref{eq:eq_rho_h2}) is equal to Eq. (\ref{eq:eq_rho_h}) if $w\approx-1$. Hence, in this scenario the chemical potential should be approximately zero.

Eqs. (\ref{eq:n_de}) and (\ref{eq:n_de3}) do not include the case when $\beta\rightarrow 0$. In this limit both particles and holes are equally distributed , so that $n_p=n_h=n_{\max}/2$. The full results are then
\begin{align}
n_p  &= \frac{Cg}{6\pi^2}\beta \mathscr{F}^{u_{\max},\,+\beta}_{u_{\min}, -2}+\frac{n_{\max}}{2}\label{eq:n_p_f}\,,\\
    n_h &= \frac{Cg}{6\pi^2}|\beta| \mathscr{F}^{u_{\max},\,-\beta}_{u_{\min}, -2}+\frac{n_{\max}}{2}\label{eq:n_h_f}\,,
    \end{align}
    where the subscripts $\pm \beta$ in $\mathscr{F}^{u_{\max}}_{u_{\min}, -2}$ are simply to indicate the regime in which the integrals are referred to.
    
\begin{figure}[h]
    \centering
    \includegraphics[scale=0.5]{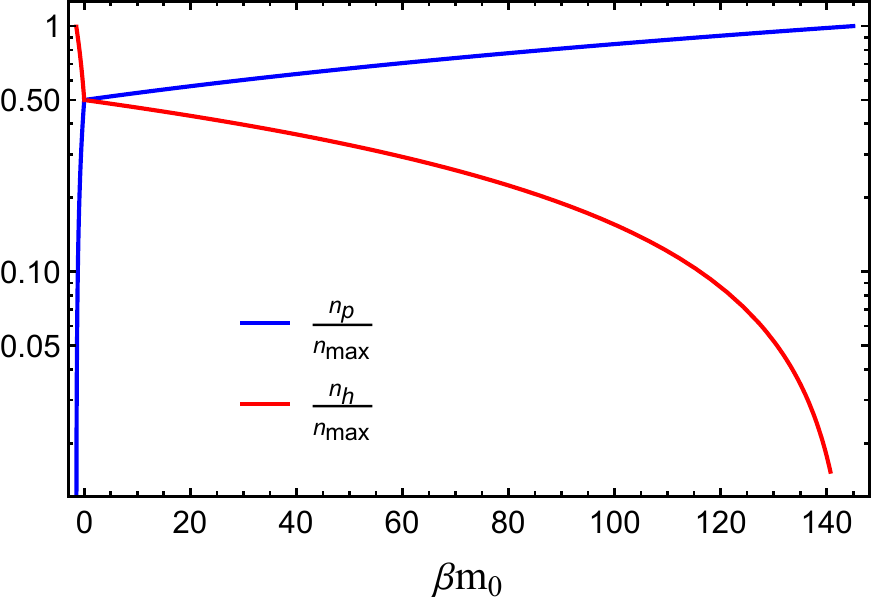}
    \caption{Ratio between the particle (hole) number density to the maximum number density in blue (red), as a function of $\beta m_0$. The temperature increases from right to left. }
    \label{fig:F_10}
\end{figure}

    Using Eqs. (\ref{eq:n_p_f}) and (\ref{eq:n_h_f}) we plot in Fig. \ref{fig:F_10} the ratios $n_p/n_{\max}=\beta m_0 \mathscr{F}^{u_{\max},\,+\beta}_{u_{\min}, -2}+1/2$ and $n_h/n_{\max}=|\beta| m_0 \mathscr{F}^{u_{\max},\,-\beta}_{u_{\min}, -2}+1/2$ with $u_{\min}=3$, as a function of $\beta m_0$. The maximum combination $\beta m_0$ for positive temperature that gives $n_p=n_{\max}$ is $\beta m_0=145$, while $n_h=n_{\max}$ for $\beta m_0=-1.5$. Soon after the FDE particles reach the transition $+T\rightarrow -T$, they quickly reach the maximum negative temperature $T=0^-$. This behavior happens because both energy and temperature increase with the volume, then when the energy of the particles reach the cutoff the volume continue  increasing and  $\epsilon/T$ is no longer constant, enabling a transition from positive to negative temperatures  $\beta\Lambda\ll 1$. However, once the particles reach the cutoff, they  populate the highest energy state, and the number density and energy density   of the system are at their corresponding maximum values. Therefore, the transition $+\infty\rightarrow -T\rightarrow 0^-$ happens  in a very short time interval. Different values of $u_{\min}$ lead to similar behavior. If $u_{\min}=10$, then  $n_p=n_{\max}$ for $\beta m_0 =10^6$, and $n_h=n_{\max}$ for $\beta m_0=-5$. The maximum $+\beta$ for a given mass is the minimum possible temperature at the transition of canonical nonrelativistic matter to noncanonical nonrelativisitic matter, i.e., when the noncanonical kinetic term dominates and the temperature of the system stops decreasing to start increasing. 
    
    Since  positive temperatures scale with the volume, a variation from $\beta m_0$ to zero corresponds to a decrease of $\beta m_0(z_{\beta=0}+1)^3/(z_{*}+1)^3$, where the redshift $z_*$ corresponds when $n_p=n_{\max}$, that is, at the beginning of the FDE evolution.   If $\beta m_0=10^6 $ and the  redshift of FDE formation is $z_*=10^{12}$, at redshift $10^8$ the combination $\beta m_0$ is reduced to $10^{-6}$. A similar reduction happens for $z_*=10^9$ and  $z_{\beta=0}=10^5$. For the smaller value $\beta m_0 =145$ the transition from positive to negative temperatures happens much faster. 
    
    The fact that $n_h\approx n_{\max}\gg n_p$, soon after the temperature becomes negative, indicates from the number conservation equation that $\dot{n}_h=0$, thus the holes are not diluted as the Universe expands, but fill all the high-energy states. The behavior of FDE particles therefore does not suffer from the problems presented in \cite{Vieira:2016lyj} when the number of particles is conserved.

 For positive $\beta$ we have, for example, $\mathscr{F}^{10^4}_{10,-1}=10^{-6}$ while $\mathscr{F}^{10^4}_{10,-1}=7$ for negative $\beta$. If $u_{\min}$ is one order of magnitude smaller, then $\mathscr{F}^{10^4}_{3,-1}=10^{-2}$ for positive $\beta$  and $\mathscr{F}^{10^4}_{3,-1}=8$ for negative  $\beta$. These results indicate that  the constant $C$ should be of order $10^{-46}$ GeV$^4$, i.e., $C=\lambda^6 M_{Pl}^{-2}$, with $\lambda\sim 10^{-8/6}$ GeV, to give the observed value of the vacuum energy. Since the maximum energy density depends logarithmically  on the cutoff scale, even arbitrarily large values of $\Lambda\sim 10^{19}$ GeV would still give the observed value of the vacuum energy if the constant is $C\sim 10^{-47}$ GeV$^4$. Using the parametrization including the Planck mass, the length scale is $(M_{Pl}^2/\lambda^6)^{1/4}$, indicating that the length $\lambda^{-1}$ is of order $10^{7.8}l_p^{1/3}\sim 10^{3/2}$ GeV$^{-1}$, where $l_p$ is the Planck length.

\begin{figure}
    \centering
    \includegraphics[scale=0.19]{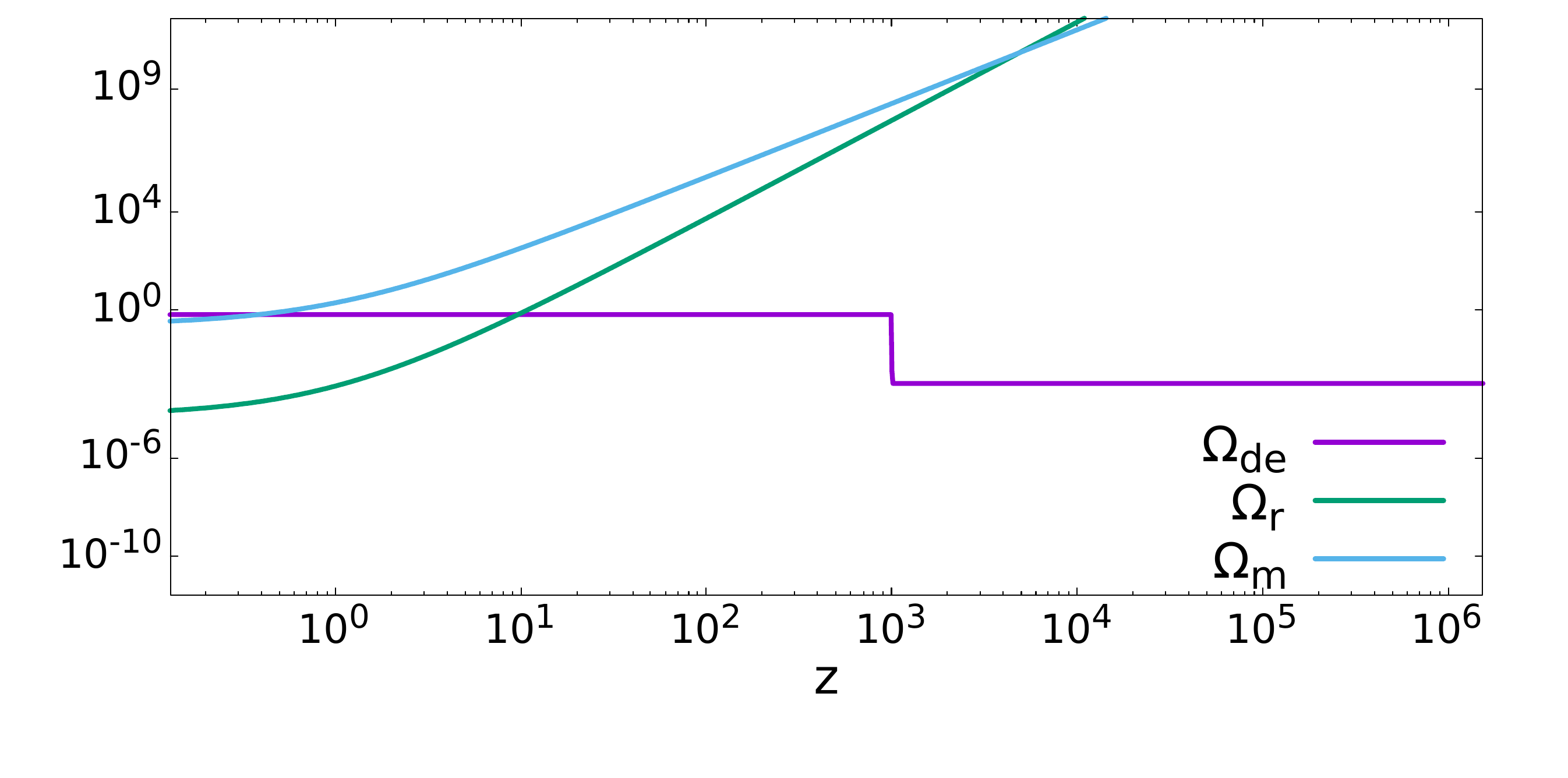}
    \caption{Evolution of  the density parameter for radiation, matter and dark energy with a illustrative transition from $+T$ to $-T$ at redshift $10^3$.}
    \label{fig:rho}
\end{figure}

The energy density is constant for both particles and holes, and we can see its evolution and the transition from $+T$ to $-T$ in Fig. \ref{fig:rho}, where we selected the corresponding initial energy $\beta \epsilon_i=3$, which is the minimum allowed value for this parameter. Other choices for the minimum energy would give very similar results. Since the equation of state should be minus one, DE perturbations are practically zero.

In order to investigate the impact of this transition in the early Universe, we will use cosmological observations to constrain the redshift of the transition $z_{\beta=0}$ and the usual six $\Lambda$CDM parameters. We take the same initial energy as before and use an adapted version of {\tt CLASS} \cite{blas2011cosmic}, along with {\tt MontePython} \cite{Audren:2012wb,Brinckmann:2018cvx} to constrain the cosmological parameters. 

We use different set of observations from the following surveys: CMB anisotropy results from   \textit{Planck} high-$\ell$ and low-$\ell$ temperature and polarization power spectra (TT, TE, EE) \cite{Planck:2019nip}  and lensing measurements \cite{Planck:2018lbu}; BAO measurements from 6dFGS \cite{beutler20116df}, MGS \cite{Ross:2014qpa}, BOSS DR12 \cite{BOSS:2016wmc}, and the auto and cross-correlation of Ly$\alpha$ absorption and
quasars in eBOSS DR14 \cite{Blomqvist:2019rah,deSainteAgathe:2019voe}; and 1048 SNe from the Pantheon sample \cite{Scolnic:2017caz}. We constrain the nuisance parameters along with the cosmological ones and we use the Gelman-Rubin criterion \cite{gelman1992inference} $R-1<0.06$ to assume that the chains converged. We use a prior on the redshift  $z_{\beta=0}$ [1, $10^{10}$] so that DE dominates at very late times.

\begin{figure*}
    \centering
    \includegraphics[scale=0.4]{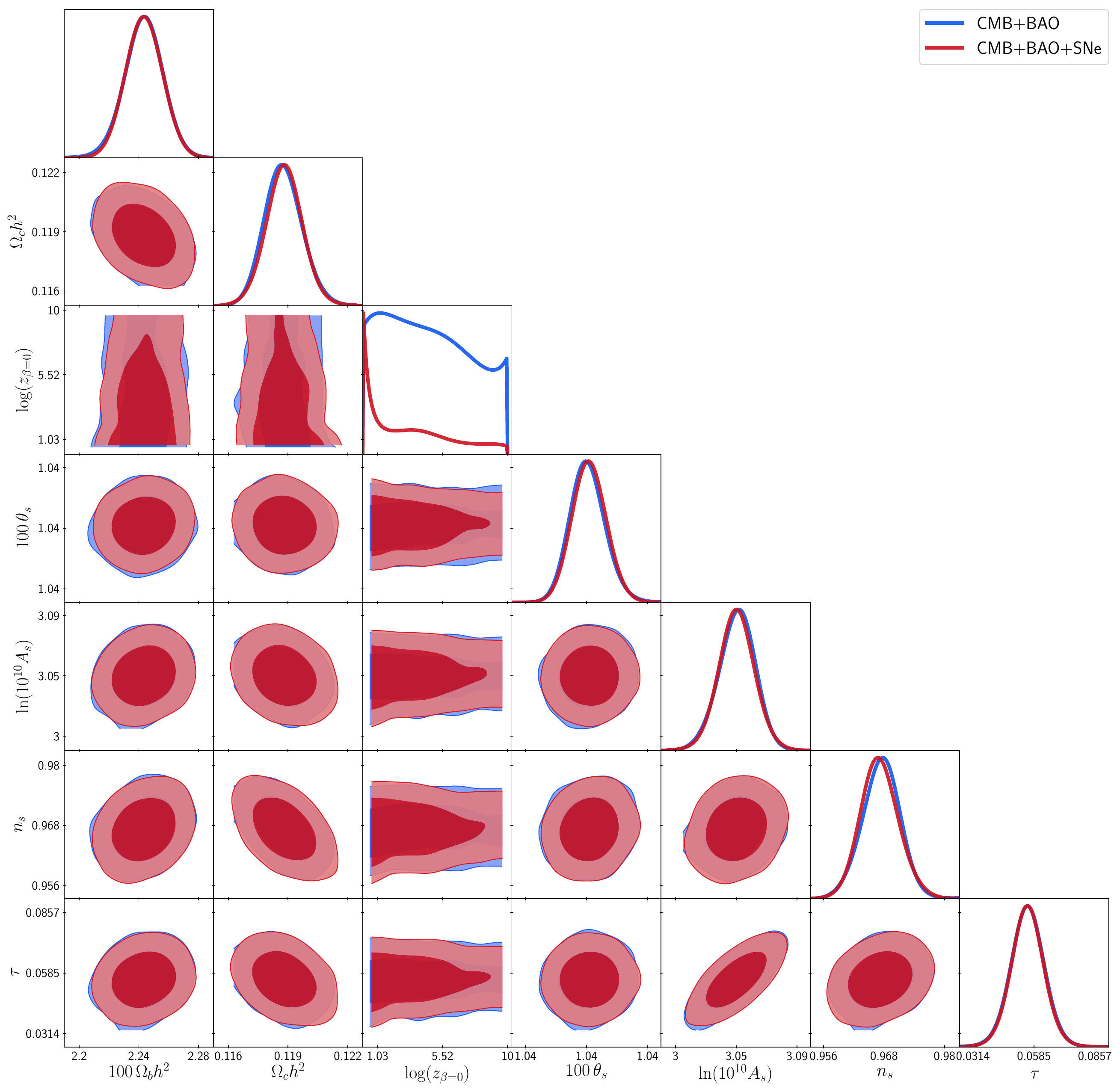}
    \caption{1D and 2D (68\% and 95\% C.L.)
posterior distributions of the cosmological parameters for the combinations CMB + BAO and CMB + BAO + SNe.}
    \label{fig:triangle}
\end{figure*}

We show in Fig. \ref{fig:triangle} the constraints on the redshift  of the transition from $+T$ to $-T$ and the other six parameters for the joint analysis of CMB + BAO and CMB + BAO + SNe. We present  in Table \ref{tab:results} the best fit and 68\% C.L. values for all parameters. The results are consistent with the $\Lambda$CDM model and the transition from positive to negative temperatures could have happened at any redshift $z_{\beta=0}>1$, thus in agreement with the cosmological observations. This agreement reflects the fact that the energy density for DE is much smaller than the corresponding ones for matter and radiation, at higher redshifts.

\begin{table*}
\caption{Cosmological parameters using CMB, BAO and SNe data.}
\label{tab:results}
\begin{tabular}{l l l l l } 
 \hline \hline
     \multirow{2}{*}{Parameter}      &       \multicolumn{2}{c}{CMB + BAO}             &           \multicolumn{2}{c}{CMB + BAO + SNe}                \\
          \cline{2-5}
  & Best fit & 68\% limits & Best fit & 68\% limits \\ \hline 
$100\,\Omega{}_{b }h^2$ &$2.246$ & $2.244_{-0.014}^{+0.015}$ & $2.243$ & $2.244_{-0.015}^{+0.014}$\\ 
$\Omega{}_{c }h^2$ &$0.1189$ & $0.119_{-0.001}^{+0.001}$ &$0.1189$ & $0.119_{-0.001}^{+0.001}$  \\ 
$100\,\theta{}_{s }$ &$1.0421$ & $1.0420_{-0.0003}^{+0.0003}$  &$1.0419$ & $1.0420_{-0.0003}^{+0.0003}$ \\ 
$\ln(10^{10}A_{{s} })$ &$3.046$ & $3.045_{-0.015}^{+0.015}$ &$3.034$ & $3.045_{-0.015}^{+0.015}$ \\ 
$n_{s }$ &$0.9669$ & $0.9674_{-0.0040}^{+0.0042}$  &$0.9658$ & $0.9672_{-0.0040}^{+0.0040}$  \\ 
$\tau$ &$0.0541$ & $0.0553_{-0.0075}^{+0.0078}$ &$0.0509$ & $0.0556_{-0.0079}^{+0.0077}$  \\ 
$\log(z_{\beta=0})$ &$4.48$ & $4.63_{-4.44}^{+1.54}$ &$3.64$ & $3.80_{-3.54}^{+1.13}$  \\ 
\hline
$H_{0 }$ &$67.93$ & $67.86_{-0.57}^{+0.49}$  &$67.83$ & $68.00_{-0.68}^{+0.37}$ \\ 
$\Omega{}_{m }$ &$0.3078$ & $0.3090_{-0.0061}^{+0.0066}$ &$0.3086$ & $0.3079_{-0.0053}^{+0.0069}$  \\ 
\hline  
$\chi^2_{\min}$ &  \multicolumn{2}{c}{ $2790$}             &           \multicolumn{2}{c}{$ 3817$}                \\ 
\hline \hline
\end{tabular}
\end{table*}

\section{Conclusions} \label{sec:conclu}
In this paper we have investigated further aspects of the FDE model. First, we analyzed FDE with a nonzero chemical potential. In this case,  FDE possesses a phantom behavior and the equation of state parameter has a lower limit. The combination of temperature and chemical potential $\beta \mu$ is constrained by  measurements of $w$ by \textit{Planck}, giving an upper bound of  $\beta\mu \lesssim 5\% \beta \epsilon$. 

In the second part of this paper we analyzed  the NAT behavior for FDE. In order to avoid a divergence in the energy when the momentum goes to zero, a cutoff is introduced. This cutoff is precisely a requirement for NAT, thus indicating a connection between FDE and NAT, if FDE is composed of fermions. In this case the equation of state parameter should be equal to minus one and the chemical potential should be zero. We have shown that when the energy of the system reaches the cutoff, the transition to NAT happens and in fact, all high-energy states become quickly populated, leading in turn to the maximum energy density and number density for the system. Negative temperatures saturates ($T=0^-$) in a very short time and the holes are no longer diluted as the Universe expands. Using combined data from CMB, BAO and SNe, we investigated  the redshift of the transition from $+T$ to $-T$ and the results  indicated that this transition is allowed at any redshift larger than one. This connection between FDE and NAT elucidates the properties of the present model and shows a possible fate of the FDE gas.

\begin{acknowledgements}
We thank the Alexander von Humboldt Foundation and  CAPES (process 88881.162206/2017-01) for the financial support.
\end{acknowledgements}

\bibliographystyle{unsrt}
\bibliography{main}\end{document}